\documentclass[a4paper,11pt]{article}
\usepackage[utf8]{inputenc}
\usepackage{fullpage}
\usepackage{amsmath,amssymb,amsfonts}
\usepackage{authblk}

\newtheorem{definition}{Definition}

\usepackage{tikz}
\usetikzlibrary{chains}
\usetikzlibrary{fit}
\usetikzlibrary{arrows}
\makeatletter
\pgfdeclareshape{datastore}{
  \inheritsavedanchors[from=rectangle]
  \inheritanchorborder[from=rectangle]
  \inheritanchor[from=rectangle]{center}
  \inheritanchor[from=rectangle]{base}
  \inheritanchor[from=rectangle]{north}
  \inheritanchor[from=rectangle]{north east}
  \inheritanchor[from=rectangle]{east}
  \inheritanchor[from=rectangle]{south east}
  \inheritanchor[from=rectangle]{south}
  \inheritanchor[from=rectangle]{south west}
  \inheritanchor[from=rectangle]{west}
  \inheritanchor[from=rectangle]{north west}
  \backgroundpath{
    \southwest \pgf@xa=\pgf@x \pgf@ya=\pgf@y
    \northeast \pgf@xb=\pgf@x \pgf@yb=\pgf@y
    \pgfpathmoveto{\pgfpoint{\pgf@xa}{\pgf@ya}}
    \pgfpathlineto{\pgfpoint{\pgf@xb}{\pgf@ya}}
    \pgfpathmoveto{\pgfpoint{\pgf@xa}{\pgf@yb}}
    \pgfpathlineto{\pgfpoint{\pgf@xb}{\pgf@yb}}
 }
}
\makeatother

\usepackage{float}
\usepackage{hyperref}			
\hypersetup{
	colorlinks=true,
	linkcolor=blue,
	urlcolor=red,
	linktoc=all
}

\usepackage{outline}

\usepackage{xargs}              
\usepackage[colorinlistoftodos,prependcaption]{todonotes}
\newcommandx{\unsure}[2][1=]{\todo[linecolor=red,backgroundcolor=red!25,bordercolor=red,#1]{#2}}
\newcommandx{\change}[2][1=]{\todo[linecolor=blue,backgroundcolor=blue!25,bordercolor=blue,#1]{#2}}
\newcommandx{\info}[2][1=]{\todo[linecolor=OliveGreen,backgroundcolor=OliveGreen!25,bordercolor=OliveGreen,#1]{#2}}
\newcommandx{\improvement}[2][1=]{\todo[linecolor=Plum,backgroundcolor=Plum!25,bordercolor=Plum,#1]{#2}}
\usepackage{subfigure}

\begin{document}

\title{Verifiable Manufacturing Using Blockchain}

\author[1]{Michael Chiu\footnote{chiu@cs.toronto.edu}}
\affil[1]{Department of Computer Science, University of Toronto, ON M5S 2E4, Canada.}
\author[2]{Jyotiraditya Panda\footnote{jyoti.panda@mail.utoronto.ca}}
\affil[2]{ Division of Engineering Science, University of Toronto, ON M5S 2E4, Canada.}
\author[3]{Abraham Goldsmith\footnote{goldsmith@merl.com}}
\affil[3]{Mitsubishi Electric Research Laboratories, Cambridge, MA 02139, USA}
\author[3]{Uro\v{s} V.~Kalabi\'{c}\footnote{kalabic@merl.com}}

\date{December 31, 2021}

\maketitle

\begin{abstract}
We propose a blockchain-based solution for enabling verifiability of manufacturing processes. We base our solution on the methodology of verifiable computing which, originally developed for cloud computing, enables clients to outsource computations to more powerful servers without the need to trust that the server correctly performed desired computation. Verifiable computing accomplishes this by enabling the client to generate cryptographic objects that the server must use to produce a cryptographic proof that verifies the correctness of results.
The black box nature of servers in cloud computing is analogous to that of the manufacturing processes of an upstream manufacturer. In this work, we develop a one-to-one correspondence between physical processes and their digital representations as state sequences which is needed for the implementation of verifiable computing. Because direct application of verifiable computing in this case would be computationally prohibitive, we introduce a blockchain to provide a computationally feasible methodology for verifiable computing applied to physical processes. We implement and show the results of our implementation on a proof of concept, developed on Hyperledger Fabric.
\end{abstract} 

%
%

\section{Introduction}
The industry-wide smart manufacturing initiative known as Industry 4.0, seeks to unify the physical with the digital not only within the factory but also the supporting supply chains outside the factory. This is generally being accomplished through the adoption of technologies such as the Industrial Internet-of-Things (IIoT), embedded devices, cyber-physical systems, Internet Protocol (IP) technologies, etc.~\cite{industry40-survey-applications-research}. A core goal of Industry 4.0 is to create interconnected systems that enable more dynamic information and work flows in contrast to traditional supply chains, with static, one-way flows. 

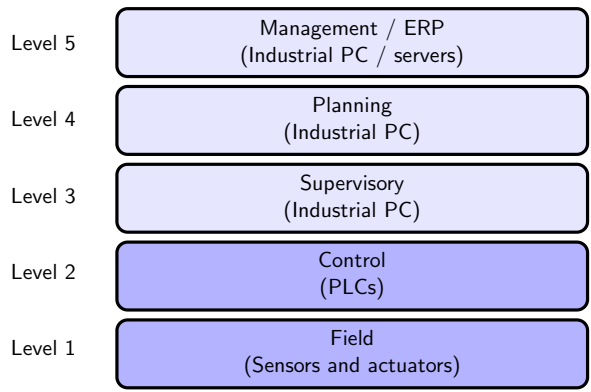
\begin{figure}[b]
\centering
\begin{tikzpicture}[
  scale=0.75,
  start chain=1 going below, 
  start chain=2 going right,
  node distance=1mm,
  desc/.style={
    scale=0.75,
    on chain=2,
    rectangle,
    rounded corners,
    draw=black, 
    very thick,
    text centered,
    text width=8cm,
    minimum height=12mm,
    fill=blue!30
    },
  it/.style={
    fill=blue!10
  },
  level/.style={
    scale=0.75,
    on chain=1,
    minimum height=12mm,
    text width=2cm,
    text centered
  },
  every node/.style={font=\sffamily}
]

\node [level] (Level 5) {Level 5};
\node [level] (Level 4) {Level 4};
\node [level] (Level 3) {Level 3};
\node [level] (Level 2) {Level 2};
\node [level] (Level 1) {Level 1};

\chainin (Level 5); 
\node [desc, it] (Archives) {Management / ERP \\ (Industrial PC / servers)};
\node [desc, it, continue chain=going below] (ERP) {Planning \\ (Industrial PC)};
\node [desc, it] (Supervisory) {Supervisory \\ (Industrial PC)};
\node [desc] (Control) {Control \\ (PLCs)};
\node [desc] (Field) {Field \\ (Sensors and actuators)};
\end{tikzpicture}
\caption{Hierarchy of industrial automation}
\label{fig:hierarchy-industrial-automation}
\end{figure}

A prerequisite in achieving the goals of Industry 4.0 is the digitization of the factory, for which there exist various proposals (such as Manufacturing Execution Systems or Enterprise Resource Planning systems). However, while digitizing the factory is necessary, it is not sufficient 
because dynamic interconnected systems, especially between multiple parties, require mutual trust. Current solutions to digitization for factory automation, to the best of the authors' knowledge, focus on a single organizational settings. Moreover, many of the solutions seek to provide or increase trust within Levels 3-5 of the hierarchy of industrial automation (see Fig.~\ref{fig:hierarchy-industrial-automation}), where the available computational resources are that of the cloud. 
However, ensuring trust at the lower Levels 1 and 2 is more beneficial to the overall hierarchy because that is where 
actual physical transformations are applied to manufactured items. However, the lower levels of 
are characterized by computationally-constrained embedded devices where, in many cases, logging is the greatest extent to which trust can be increased by current solutions. 


The desire for trust in the manufacturing process is similar to the desire for trust in outsourced computations carried out by third-party cloud service providers. 
Currently, one must fully trust that the third party 
properly carries out the computation \cite{walfish2015verifying} and this is not ideal for a number of reasons. For example, even if the third party service provider is completely trustworthy, there is a non-zero probability that the service provider may be compromised. A less malicious and more likely possibility is the introduction of errors into the outsourced computation. In both cases, one cannot definitively show that the computation was carried out correctly. Verifiable computing seeks to address this by providing a cryptography-based methodology that enables the third party to provide, along with the computation itself, a cryptographic proof that the computation was indeed correctly computed. This way, the originator of the computation need only trust the cryptographic machinery behind the generation of the proof. Importantly, the computational work required for verifying a computation should be much smaller than the actual computation \cite{gennaro2010non}.

At a high level, verifiable computing is typically concerned with the following interaction between three parties. 
A \emph{client} specifies the function $f$ to be computed and provides the input data, denoted $x$. A \emph{server} computes $f(x)$. Finally, a \emph{verifier} checks the correctness of the result, i.e., that $y = f(x)$. A verifiable computing scheme enables clients to generate the cryptographic objects that both the server and the verifier use to convince the client that the computation was carried out correctly.
In typical settings considered in verifiable computing, the function $f$ and the input $x$ to be verified differ with every instance of the problem. This reflects the typical workload encountered in a cloud computing setting, i.e., many singular computations from different users. 
Workloads found in a manufacturing setting, however, consist of repeated applications of the same function, i.e., some physical transformation $f$, to many identical but uniquely identifiable instances of an input $x_i$, where $i$ is within a finite set of possibilities. A direct application of verifiable computing to the manufacturing setting is infeasible since a cryptographic proof would be generated for each product on a manufacturing line. Each proof, while not identical, due to the entropy, would be functionally redundant since each proof verifies the same function on the same input. Generating verifiable proofs for each product with identical specifications is computationally wasteful, adding possibly intolerable latency to the manufacturing process and  unnecessarily increasing the storage requirements by retaining a redundant proof for each manufactured product. 

In this work, we introduce a 
blockchain-based verifiable computing system for manufacturing processes. Two main components comprise the system: verifiability at the field device level and a permissioned blockchain network that reduces the computational cost of verification at the field device level.
Verifiability at the field device level is achieved by ensuring verifiability of controllers, i.e., devices in the control level of the hierarchy of industrial automation. Blockchain is used to enable the use of a single proof for verifying all items 
manufactured according to the same specifications.
We present an application of the idea in the form of a proof ofconcept (PoC). We use Hyperledger Fabric \cite{hlf2018}, a Go-based framework for permissioned blockchains, to implement the blockchain network. For the verifiable computing component, we use gnark, a Go-based library that provides an API-based approach to generating polynomial circuits. 

The literature has explored many systems that implement verifiable computing including those that are based on secure hardware \cite{parno2011bootstrapping}, verifying specific functions \cite{golle2001uncheatable}, auditing \cite{belenkiy2008incentivizing}, and cryptography and complexity theory \cite{parno2013pinocchio}. 
The topic of applying blockchain to industrial systems is covered in \cite{alladi2019blockchain,blockchain-embedded-system-accountability,liang2019secure,zhuang2020blockchain}. To the best of the authors' knowledge, the literature has not explored applying verifiable computing to manufacturing processes.

The rest of this paper is structured as follows. Section~\ref{sec:verifiable-field-devices} discusses a theoretical framework for verifiable manufacturing. Section~\ref{sec:implementation} details the implementation of our framework in a PoC. Section~\ref{sec:concl} is the conclusion.

%
%

\section{Verifiable Manufacturing}
\label{sec:verifiable-field-devices}

Verifiable computing provides a 
cryptography-based framework for ensuring that, given a computation $f$ on some data $x$, an accompanying verifiable proof \footnote{Also known as a transcript} $\sigma_y$ can be generated by the server such that any third party can use $\sigma_y$ to verify that $y = f(x)$. In this section, we explain how the verifiable computing framework, originally developed for  cloud computing, 
can be adapted and applied to industrial manufacturing processes. 

Verifiability of manufacturing processes is desirable for many of the same reasons as in cloud computing. It is often the case that manufacturing companies have no visibility into the manufacturing processes employed by their suppliers. Suppliers are essentially black boxes that provide a product that conforms to a specification. A similar situation exists in cloud computing, where clients have no visibility into the operations of the servers responsible for their outsourced computations. Errors in 
industrial automation typically have an outsize impact as they lead to physical, as opposed to software, faults. 
For example, small errors in configuration, nearly impossible to detect, can be extremely costly for downstream OEMs (original equipment manufacturers);
for this reason, downstream OEMs typically expend additional effort to test upstream products, in order to ensure specifications. %


To introduce verification of physical manufacturing processes, 
we introduce verifiability at Levels 1 and 2 of the hierarchy of industrial automation (see Fig.~\ref{fig:hierarchy-industrial-automation}), 
since it is within these layers where actual physical transformations are carried out. 
At these levels, devices are usually computationally constrained, and it is difficult to apply verifiable computing directly. 
In the rest of the section, we show how verifiable computing can be applied in the industrial context. 

We begin by defining a verifiable computing scheme, after which we show how that scheme can be applied to control and field devices, and finally we discuss how a permissioned blockchain can be used to make verifiable proofs practical in a factory automation setting. 
 




\subsection{Verifiable Computing}
\label{subsec:verifiable_computation}

We begin with the following definition.
\begin{definition}[\cite{demirel2017privately}]
A \emph{verifiable computing scheme} consists of the following (probabilistic, polynomial-time) 
algorithms:
\begin{enumerate}
  \item $\textrm{KeyGen}(1^\lambda,f) \rightarrow (\textsf{sk,vk,ek})$: a setup algorithm that, given a security parameter $\lambda$ and a specification of the function $f$, generates a secret key $\textsf{sk}$, a verification key $\textsf{vk}$, and a public evaluation key $\textsf{ek}$ encoding $f$.
  \item $\textrm{ProbGen}(\textsf{sk},x) \rightarrow (\sigma_x,\rho_x)$: encodes the problem data by taking the data $x$ and the secret key $\textsf{sk}$ and returns the encoded data $\sigma_x$ and the decoding value $\rho_x$.
  \item $\textrm{Compute}(\textsf{ek},\sigma_x) \rightarrow \sigma_y$: computes a proof $\sigma_y$, an encoded version of the output $y=f(x)$ using the evaluation key $\textsf{ek}$.
  \item $\textrm{Verify}(\textsf{vk},\rho_x,\sigma_y) \rightarrow \{y,\bot\}$: given the verification key $\textsf{vk}$, a decoding value of the data $\rho_x$, and the encoded output (verifiable proof) $\sigma_y$, the verification returns $\bot$ if $y \ne f(x)$ and returns $y$ if $y=f(x)$.
\end{enumerate}
\label{def:verifiable_computing_scheme}
\end{definition}

\begin{figure}[tp]
\centering
\begin{tikzpicture}[
  font=\sffamily,
  every matrix/.style={ampersand replacement=\&,column sep=2cm,row sep=2cm},
  source/.style={draw,thick,rounded corners,fill=blue!20,inner sep=.3cm},
  process/.style={draw,thick,circle,fill=blue!20},
  client/.style={draw,thick,circle,fill=yellow!20},
  ans/.style={draw,thick,circle,fill=red!20},
  sink/.style={source,fill=green!20,minimum width=3cm},
  datastore/.style={draw,very thick,shape=datastore,inner sep=.3cm},
  clear/.style={},
  textbox/.style={draw,thin},
  to/.style={->,>=stealth',shorten >=1pt,semithick,font=\sffamily\footnotesize},
  every node/.style={align=center}]

\node[textbox] (verifier) at (-4,-4) {Verifier};
\node[draw=none,fill=none,align=left] (verifierlabel) at (3,0) {};
\node[clear] (v_f) at (-3,2) {$f$};
\node[sink] (keygen) at (0,2) {Keygen};
\node[ans] (v_sk) at (-1.5,0) {\textsf{sk}};
\node[ans] (v_vk) at (0,0) {\textsf{vk}};
\node[ans] (v_ek) at (1.5,0) {\textsf{ek}};
\node[sink] (probgen) at (-1,-2) {ProbGen};
\node[clear] (v_sigma_x) at (-2,-3.5) {$\boldsymbol\sigma_x$};
\node[clear] (v_rho_x) at (0,-3.5) {$\boldsymbol\rho_x$};
\node[clear] (v_x) at (-4,-1) {$x$};

\draw[to] (v_f.east) -- node[midway,above] {(1)} (keygen.west);
\draw[to] (keygen.south) -- (v_sk.north);
\draw[to] (keygen.south) -- (v_vk.north);
\draw[to] (keygen.south) -- (v_ek.north);
\draw[to] (v_sk.south) -- ([xshift=-0.5cm]probgen.north);
\draw[to] (probgen.south) -- (v_sigma_x.north);
\draw[to] (probgen.south) -- (v_rho_x.north);
\draw[to] (v_x) -- (-4,-2) -- node[midway,left,xshift=-0.7cm,yshift=-0.15cm] {(2)} (probgen.west);

\node[textbox] (prover) at (-4,-7) {Prover};
\node[draw=none,fill=none,align=left] (proverlabel2) at (3,-7) {};
\node[sink] (compute) at (-1,-6) {Compute};
\node[clear] (p_sigma_y) at (2.5,-6) {$\boldsymbol\sigma_y$};
\node[draw=none,fill=none,align=left] (x_sigma_y_label) at (2.5,-3.5) {};

\draw[to] (v_f) -- node[midway,left,yshift=3cm] {(3a)} (-3,-6) -- (compute.west);
\draw[to] (v_ek) -- node[midway,right] {(3c)} (1.5,-3.5) -- ([xshift=0.8cm]compute.north);
\draw[to] (v_sigma_x.south) -- node[midway,right,yshift=0.4cm] {(3b)} ([xshift=-1cm]compute.north);

\draw[to] (compute) -- node[midway,below] {(3d)} (p_sigma_y);
\draw[to] (p_sigma_y.north) -- node[midway,right,yshift=0.5cm] {(4)} (x_sigma_y_label);

\tikzset{blue dotted/.style={draw=blue!50!white, line width=2pt,
                            dash pattern=on 1pt off 4pt on 6pt off 4pt,
                            inner sep=4mm, rectangle, rounded corners}};

\tikzset{red dotted/.style={draw=red!50!white, line width=1pt,
                        dash pattern=on 1pt off 4pt on 6pt off 4pt,
                        inner sep=4mm, rectangle, rounded corners}};

\node (first dotted box) [red dotted,
                        fit = (compute) (proverlabel2) (prover)] {};
                        
\node (first dotted box) [blue dotted,
                        fit = (verifier) (keygen) (v_sigma_x) (verifierlabel) (v_f)] {};

\end{tikzpicture}
\caption{Non-interactive verifiable computing performed by a prover and a verifier}
\label{fig:general-verifiable-computing-scheme}
\end{figure}
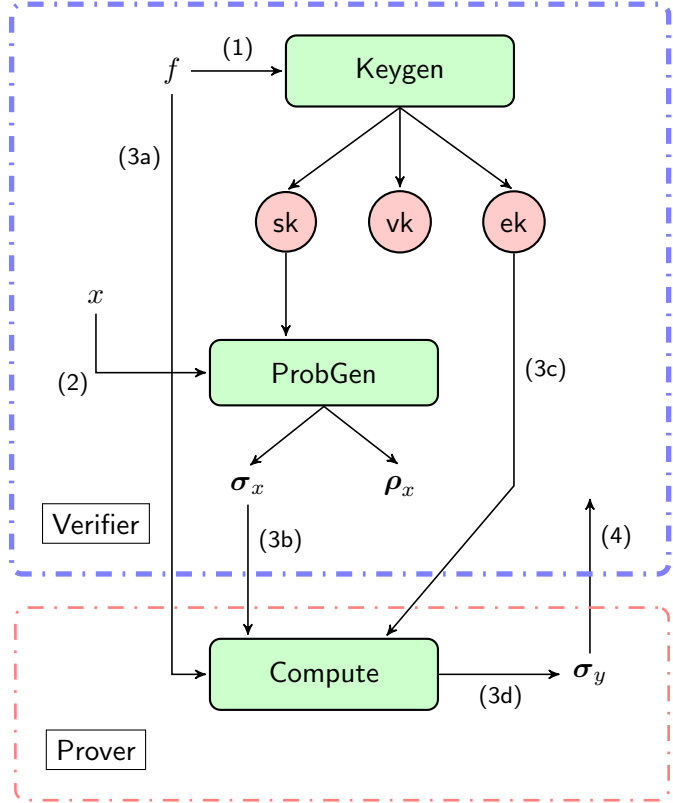

A verifiable computing scheme enables a client to outsource computations to an untrusted server. The overall process is illustrated in Fig.~\ref{fig:general-verifiable-computing-scheme}. The prover first 1) generates the necessary cryptographic keys by executing \textsf{KeyGen}$(1^\lambda,f)$, which returns the secret (sk), verification (vk), and evaluation (ek) keys. Next, the prover encodes the problem data by 2) running \textsf{ProbGen}$(\textsf{sk},x)$ using the secret key \textsf{sk} and returns the encoded problem data $\sigma_x$ and the decoding value $\rho_x$. The verifier then sends the function 3a), the encoded problem data 3b), and the evaluation key 3c) to the prover who computes the desired computation $f(x)$ and executes \textsf{Compute}(\textsf{ek},$\sigma_x$) returning the encoded output $\sigma_y$ 3d) that constitutes the cryptographic proof, which 4) is sent to the verifier. Note that a verifiable computing scheme is known as a ``privately verifiable computing scheme'' if $\textsf{sk} = \textsf{vk}$. Otherwise, it is known as a ``publicly verifiable computing scheme''. Publicly verifiable schemes enable a client to publish the verification key where any third party possessing $\rho_x$ and \textsf{vk} can verify the correctness of an outsourced computation.

Verifiable computing schemes can be implemented using a variety of cryptographic 
technologies such as: interactive proofs \cite{goldwasser1989knowledge}, probabilistically checkable proofs \cite{arora1998probabilistic}, fully homomorphic encryption \cite{gentry2009fully}, and argument systems \cite{brassard1988minimum}. Early work in implementing verifiable computing using the aforementioned technologies were not practical due to their high computational costs, but
recent technologies such as \cite{parno2013pinocchio} and \cite{costello2015geppetto} have made verifiable computing practical.


In practice, the most important step in implementing verifiable computing is the translation from a human-readable programming language to its 
backend equivalent. This translation process is split into a frontend and backend phase, similar to how traditional computer code is compiled from a human-readable language to an intermediate representation (IR) in the frontend and from the IR to a platform-specific form in the backend. For example, \cite{ben2013snarks} and \cite{kosba2018xjsnark} provide not only advances in implementing verifiable computing schemes but also toolchains that convert C and Java programs, respectively, into arithmetic circuits, whose details we describe later. 




\subsection{Verifiability for Physical Processes}

Devices in the field level, or ``field devices'', broadly refer to devices such as CNC (computer numerical control) machines, machining tools, actuators, valves, sensors, etc. They often have 
limited or no 
compute and are used to 
perform physical transformations on a production floor. Field devices possess a finite number of atomic operations, each of which can be modeled as a state in a finite state machine \cite{michaloski1998framework}. As such, a sequence of states, or 
``state sequence'', corresponds to a field device effecting a physical transformation, i.e., there is a one-to-one relationship between a state sequence and a physical transformation. 

Field devices are driven and coordinated by specialized industrial computers, known as PLCs (programmable logic controllers), that are designed for rugged industrial environments and for controlling electromechanical devices. PLCs are examples of devices in the control level, or ``control devices'', responsible for coordinating field devices. It is necessary for field devices to be controlled by control devices because, on a production floor, field devices need to be precisely coordinated with other field devices to effect physical transformations. Without control devices, field devices would need to make 
a connections to every other field device. Moreover, field devices do not possess the computing power to be able to coordinate with other field devices since that would require, at a minimum, a real-time enabled kernel.

For these reasons, while a state sequence is equivalent to a physical transformation, the state transitions occur on the PLC and not on the field device. As a PLC transitions through the states of a state sequence, it instructs the corresponding field device to perform an operation. Concurrently, or after the operation is completed, the field device returns information to the PLC such as feedback on the current state of the field device, whether the field device encountered an error performing the operation, the configuration of the field device, etc. The controller uses this information to determine whether or not to transition to the subsequent state. 

Thus, enabling verifiability of a manufacturing process is equivalent to ensuring that all physical transformations are made verifiable. Consequently, state sequences must be made verifiable due to their one-to-one correspondence to a physical transformation. Note that since state transitions occur on a PLC and not within a field device, a state sequence can comprise multiple physical transformations from multiple field devices. Moreover, the PLC itself must be made verifiable in order for a chain-of-trust to be established, however, this is 
outside the scope of this paper.

\subsection{State Sequence Circuits}
\label{subsec:state-circuits}

State sequences are digital representations of physical transformations within a manufacturing setting. Consequently, physical manufacturing processes can be made verifiable by making state sequences verifiable.
To this end, we begin by detailing the modeling of state sequences. 
Since field devices can be modeled as finite state machines, it follows that state sequences can be modeled as sequences of a finite number of integers,
\begin{equation*}
\textbf{s} = \{s_0,s_1, \dots, s_n\},
\label{eq:state-transition-seq}
\end{equation*}
where $s_i \in \mathbb{Z}$ for $i = 0,\dots,n$, and $n$ denotes the total number of states in the state sequence. However, representing a state sequence only as a series of integers is insufficient since, in many cases, additional information from the field device that is not inferred from its current state must also be verified. For example, it is often necessary, at the beginning of a manufacturing run, to ensure that a field device is operating with the correct configuration. Field devices, in many cases, can also return this information as a string, which can represent a configuration ID or a hash of the configuration. Note that auxiliary information is only useful at certain stages of a field device's operation. That is, auxiliary information should be coupled with the states of a field device. Thus, it is natural to define a state sequence to be a series of tuples, as in the following.
\begin{definition}
A \emph{state sequence} is a digital representation of a physical transformation directed by a PLC and is represented as a sequence of tuples,
\begin{equation*}
\textbf{s} = (s_0,a_0), \dots, (s_n,a_n)
\label{eq:state-transition-seq-tuple}
\end{equation*}
where 
$a_i$, the \emph{auxiliary data}, can be either an integer, a string, or empty for any $i = 0,\dots, n$.
\label{def:state-seq}
\end{definition}


\subsubsection*{Verifying State Sequences}

State sequences 
must be converted to arithmetic circuits in order to use the 
machinery necessary to generate verifiable proofs. Arithmetic circuits are used to 
convert basic arithmetic equations into a series of logical tests such that a value satisfying an arithmetic equation is equivalent to the logical test being satisfied \cite{walfish2015verifying}. In the manufacturing setting, downstream OEMs send the specifications of their product to their upstream manufacturers which usually generates the PLC program to be run. Therefore, downstream OEMs, in most cases, are only concerned that an agreed-upon state sequence was executed. 
To verify this, we can implement a set-membership logical test for the state sequences and auxiliary data, 
meaning that the $f$ in a verifiable computing scheme 
can consist of a series of set-membership tests for integers and strings.
For a tuple $(s_i,a_i)$, the state component, being an integer, is straightforwardly implemented in an arithmetic circuit. The only difficulty is in the 
comparisons of the auxiliary information component of a state tuple, when $a_i$ is a string. Computationally, this can be remedied by obtaining an integer representation of the string $a_i$, from which we can reduce the check for equality of strings to the equality of integers.

%
%
\subsection{Blockchain for Verifiable Proofs}
\label{subsec:blockchain-verifiable-proofs}

The main issue preventing verifiable computing from being applied to manufacturing is the prohibitive computational cost, in both compute and storage, associated with generating a verifiable proof for each manufactured item. This is because, contrary to typical workloads in cloud computing, workloads in manufacturing, i.e., within a set of items of the same specification, consist of repeated applications of a physical transformation to identical but uniquely identifiable inputs. A straightforward application of verifiable computing is impractical since it would result in the generation of a verifiable proof for every item. We note that, while each generated verifiable proof is not bit-for-bit identical, they are, however, functionally identical since they all verify that the same physical transformation was applied to the same input; the only difference between each verifiable proof is the entropy used.



To solve this problem, we introduce a blockchain solution 
enabling the reuse of a single verifiable proof for all items of the same specification.
While a verifiable proof can be reused without the use of blockchain if the verifiable proof and its associated cryptographic keys are properly distributed to participants, a verifiable proof by itself does not enable the verifiability of manufacturing processes. Assuming that the verifiable proof is in some trusted shared location or distributed to the verifier, the prover, i.e., PLC, could send a copy of the state sequence to the verifier. However, this would not implement important aspects of verification, such as 
immutable records of an attempt at verification of a physical transformation applied to a physical object and subsequent results, and these aspects are certainly possible using a blockchain solution.
Moreover, without a common network built on top of blockchain, the possibility exists that the manufacturer would be able to equivocate its results by sending different verifiable proofs to different parties. For these reasons, blockchain is a necessary component for the application of verifiable computing to physical manufacturing processes.

Our solution relies on embedding verifiable proofs within smart contracts. We denote such smart contracts as \emph{verification contracts}.
Smart contracts are programs that are stored and run on the blockchain; they have their own address and are invoked by sending transactions to their address; 
they are stored on and executed by every full node participating in the blockchain network. Combining a verifiable proof and a smart contract obviates the need to manage keys associated with a verifiable proof and the need to trust that all participants have received the same verifiable proof.

\begin{figure}[tp]
\centering
\begin{tikzpicture}[
  font=\sffamily,
  every matrix/.style={ampersand replacement=\&,column sep=2cm,row sep=2cm},
  source0/.style={draw,thick,rounded corners,fill=green!20,inner sep=.3cm},
  source/.style={draw,thick,rounded corners,fill=green!20,inner sep=.3cm},
  source2/.style={draw,thick,rounded corners,fill=green!20,inner sep=.3cm,minimum width=3.5cm},
  process/.style={draw,thick,circle,fill=blue!20},
  client/.style={draw,thick,circle,fill=yellow!20},
  sink/.style={source,fill=green!20},
  datastore/.style={draw,very thick,shape=datastore,inner sep=.3cm},
  dots/.style={gray,scale=2},
  to/.style={->,>=stealth',shorten >=1pt,semithick,font=\sffamily\footnotesize},
  every node/.style={align=center}]

\node[draw=none,fill=none,align=left] (labelnode3) at (4,-1) {};
\node[draw,align=left] at (2.7,-1.5) {Verification Contract};
\node[draw=none,fill=none,align=left] (labelnode1) at (-3,.6) {};
\node[draw=none,fill=none] (proc) at (-2,-1.4) {};
\node[source,fill=red!20] (instr) at (-2,3) {Verifiable \\ Proof };
\node[client,fill=red!20] (sysres) at (2.6,3) {Verification \\ Key };
\node[draw,align=left] at (-2,5.1) {Blockchain Node};
\node[draw=none,fill=none,align=left] (labelnode2) at (-2,5) {};


\node[source0] (cc) at (0.6,0) {Verifiable Proof \\ Logic};

\draw[to] (cc) -- node[midway,left]{Compilation} (.6,3) -- (instr);
\draw[to] (cc) -- (.6,3) -- (sysres);

                                            
\node[draw=none,fill=none,align=left] (labelnode5) at (4,3.4) {};


\tikzset{blue dotted/.style={draw=blue!50!white, line width=2pt,
                          dash pattern=on 1pt off 4pt on 6pt off 4pt,
                            inner sep=4mm, rectangle, rounded corners}};

\tikzset{red dotted/.style={draw=red!50!white, line width=1pt,
                      dash pattern=on 1pt off 4pt on 6pt off 4pt,
                        inner sep=4mm, rectangle, rounded corners}};
                        
\tikzset{cyan dotted/.style={draw=cyan!50!white, line width=1pt,
                      dash pattern=on 1pt off 4pt on 6pt off 4pt,
                        inner sep=4mm, rectangle, rounded corners}};

\node (first dotted box) [blue dotted,
                        fit = (labelnode1) (proc) (labelnode3)] {};
                        
\node (second dotted box) [red dotted,
                        fit = (instr) (labelnode2) (labelnode3) (proc) (labelnode1)] {};
                        
                        
\end{tikzpicture}
\caption{Local compilation of a verification contract}
\label{fig:verification_contract_compile}
\end{figure}
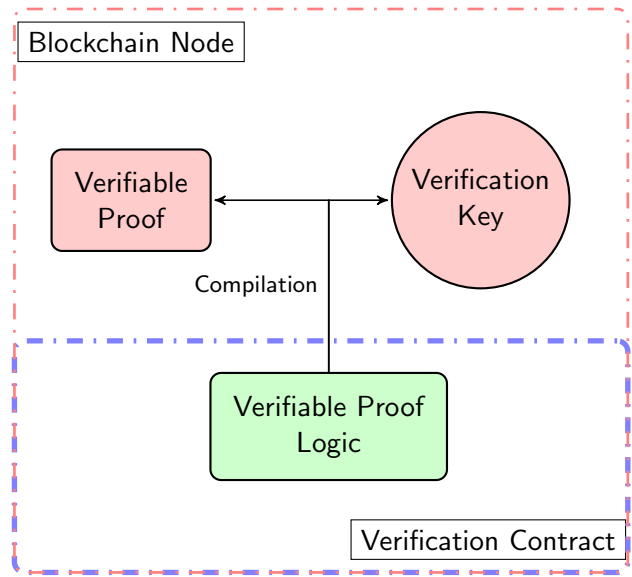

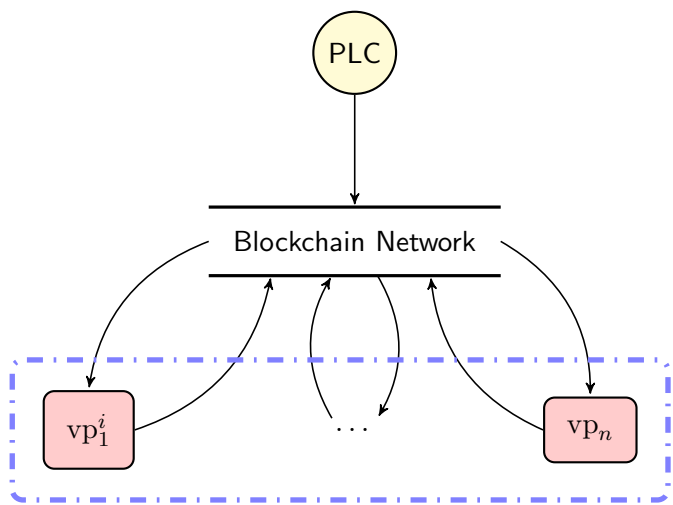
\begin{figure}[tp]
\centering
\begin{tikzpicture}[
  font=\sffamily,
  every matrix/.style={ampersand replacement=\&,column sep=2cm,row sep=2cm},
  source/.style={draw,thick,rounded corners,fill=red!20,inner sep=.3cm},
  process/.style={draw,thick,circle,fill=orange!20},
  client/.style={draw,thick,circle,fill=yellow!20},
  sink/.style={source,fill=green!20},
  datastore/.style={draw,very thick,shape=datastore,inner sep=.3cm},
  dots/.style={gray,scale=2},
  to/.style={->,>=stealth',shorten >=1pt,semithick,font=\sffamily\footnotesize},
  every node/.style={align=center},
  clear/.style={},]

\node[client] (plc) at (-0.5,2)  {PLC};
\node[source] (vp1) at (-4,-3) {$\textrm{vp}^i_1$};
\node[source] (vpn) at (2.6,-3) {$\textrm{vp}_n$};
\node[clear] (dots) at (-.5,-3) {$\dots$};

\node[datastore] (dots2) at (-0.5,-.5) {Blockchain Network};

\draw[to] (plc.south) -- (dots2.north);
\draw[to] (dots2.west) to[bend right] (vp1.north);
\draw[to] (vp1.east) to[bend right] ([xshift=-1.1cm]dots2.south);

\draw[to] ([xshift=-0.3cm]dots.north) to[bend left] ([xshift=-0.3cm]dots2.south);
\draw[to] ([xshift=0.3cm]dots2.south) to[bend left] ([xshift=0.3cm]dots.north);

\draw[to] (vpn.west) to[bend left] ([xshift=1cm]dots2.south);
\draw[to] (dots2.east) to[bend left] (vpn.north);

Define the style for the blue dotted boxes
\tikzset{blue dotted/.style={draw=blue!50!white, line width=2pt,
                            dash pattern=on 1pt off 4pt on 6pt off 4pt,
                            inner sep=4mm, rectangle, rounded corners}};


\node (first dotted box) [blue dotted,
                        fit = (vp1) (vpn)] {};
                        

\end{tikzpicture}
\caption{Process invoking a verification contract}
\label{fig:invoking_verification_compile}
\end{figure}

Figs.~\ref{fig:verification_contract_compile} and \ref{fig:invoking_verification_compile} illustrate the general scheme.
As shown in the figures, smart contracts are broadcast to the network by distributing the source code to each full node where they are checked for validity and compiled. In the case of verification contracts, a smart contract contains logic implementing a verifiable computing scheme that each full node compiles locally, as shown in Fig.~\ref{fig:verification_contract_compile}. Thus, each full node contains a verifiable proof that is functionally but not bit-for-bit identical, since different entropy was used in their generation, 
to verifiable proofs generated from the same verification contract compiled on other full nodes. 

Given the same input, the verifiable contracts on all full nodes return the same result. The verification contract is trusted throughout the blockchain network since its source code was distributed to all participants. Finally, since the verification contract is a smart contract, all attempts at verification and the subsequent results are memorialized on the immutable blockchain. This is shown in Fig.~\ref{fig:invoking_verification_compile} where a request for verification is first sent to the network and also memorialized. The nodes read the request from the blockchain and proceed to run the verification contract. The results, which must agree through consensus, are written back to the blockchain.



\subsection{Overview of Scheme}

\begin{figure}[t]
\centering
\begin{tikzpicture}[
font=\sffamily,
every matrix/.style={ampersand replacement=\&,column sep=2cm,row sep=2cm},
source/.style={draw,thick,rounded corners,fill=blue!20,inner sep=.3cm},
process/.style={draw,thick,circle,fill=blue!20},
client/.style={draw,thick,circle,fill=yellow!20},
client2/.style={draw,thick,rounded corners,fill=yellow!20,inner sep=.3cm},
sink/.style={source,fill=green!20},
sink2/.style={source,fill=red!20},
datastore/.style={draw,very thick,shape=datastore,inner sep=.3cm},
dots/.style={gray},
to/.style={->,>=stealth',shorten >=1pt,semithick,font=\sffamily\footnotesize},
every node/.style={align=center}]

\node[client2] (bc) at (0,3.5) {Blockchain};
\node[sink] (sc) at (0,1.5) {Verification Contract};
\node[source] (plc) at (0,-1) {PLC};
\node[source] (fd1) at (0,-4) {Field \\ Device};
\node[sink2] (item) at (-3,-4) { Product};
\node[dots] (fd2) at (3,-4) {...};

\draw[to] (item.east) -- node[midway,above] {(2)} (fd1.west);
\draw[to] (plc.south) -- node[midway,right] {(1)}  (fd1.north);
\draw[to] (plc.north) -- node[midway,right] {(4)} (sc.south);
\draw[to] (sc.north) -- node[midway,right] {(5)} (bc.south);
\draw[to] (fd1.east) -- node[midway,above] {(6)} (fd2.west);

\draw[to] (fd1) to[bend left=50] node[midway,above] {}
  node[midway,left] {(3)} (plc);

\end{tikzpicture}
\caption{
Physical transformation verification} 
\label{fig:field-device-verifiability-arch}
\end{figure}
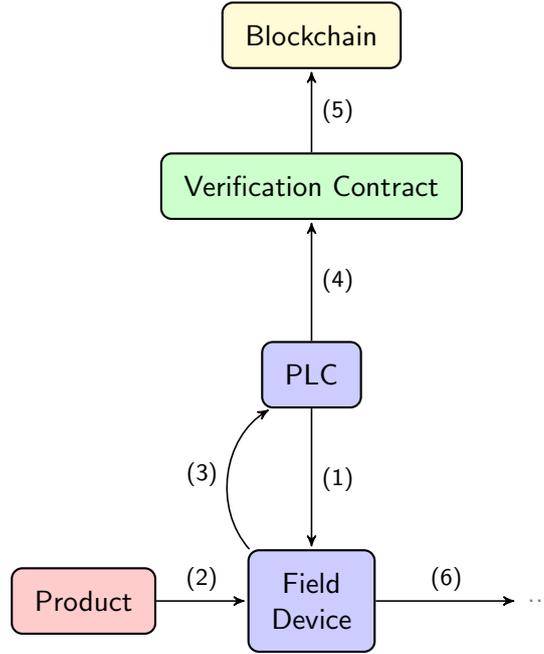

In Fig.~\ref{fig:field-device-verifiability-arch}, we present a schematic depicting how a physical transformation applied to a manufactured product is verified and how the blockchain is used. The PLC program is 1) loaded onto the PLC which begins instructing the field device. Then the product is 2) fed to the field device. The field device 3) applies the physical transformation to the manufactured product; simultaneously, the field device send feedback to the PLC on its operating state and any other auxiliary information. When the state sequence is complete, the PLC 4) forwards the state sequence to the network for verification by invoking the verification contract. Only the results of verification and a unique identifier for the manufactured item are 5) written onto the blockchain. The item then 6) progresses through the manufacturing line to the next field device.

%
%

\section{Proof of Concept}
\label{sec:implementation}

We implement a proof of concept (PoC) of a verifiable factory automation system backed by a permissioned blockchain network, where we demonstrate how field devices can be made verifiable through the verifiability of control devices. The focus of the PoC is to demonstrate the verifiability of manufacturing processes in a factory automation context by verifying the state sequences representing physical transformations and not to demonstrate the interaction between the field device and the controller. Our PoC consists of three main components: the PLC, the verifiable subsystem, and the blockchain network. We discuss the implementation of these components in the following.

\subsection{PLC}
\label{subsec:impl-rpi}

The PLC is the control unit for a series of physical transformations carried by one or more field devices. In our PoC, they are emulated using a Raspberry Pi because of their increasing use as PLC replacements \cite{bidyanath2021survey}, due to the growing virtualization of their capabilities and responsibilities \cite{alves2016virtualization}. The Raspberry Pi runs the current Ubuntu LTS (20.04) as its operating system. We used pigpio \cite{pigpio}, a C library that provides the low-level capabilities for communicating with GPIO, to develop custom code that uses GPIO to communicate to a CNC milling machine that we used as a field device.
We used the Raspberry Pi to connect to a CNC milling machine to generate sample state sequences that are representative of situations encountered in factory automation. 

\subsection{Verifiable Subsystem}

The verifiable subsystem is responsible for verifying state sequences and is implemented using smart contracts by embedding smart contracts with logic that implements a verifiable proof. In our PoC, we choose to implement a verifiable computing scheme using zk-SNARKs (Zero Knowledge Succinct Non-interactive ARguments of Knowledge) due to its many real-world implementations; zk-SNARKs are an instance of non-interactive argument systems 
and consist of three probabilistic polynomial-time algorithms, defined below.

\begin{definition}[\cite{demirel2017privately}]
A \emph{zk-SNARK} consists of the following probabilistic, polynomial-time algorithms:
\begin{enumerate}
  \item $\textsf{Setup}(1^\lambda) \rightarrow (\textsf{crs},\textsf{vrs})$: a setup algorithm that, given a security parameter $\lambda$, generates a common reference string \textsf{crs}.
  \item $\textsf{Prove}(\textsf{crs},u,w) \rightarrow \pi$: an algorithm that, given a common reference string \textsf{crs}, a statement $u$ (to be proven), and a witness $w$ (a private input) produces a proof $\pi$.
  \item $\textsf{Verify}(\textsf{vrs},u,\pi) \rightarrow \{0,1\}$: given a verification state, a statement $u$, and a proof $\pi$, either accepts or rejects the validity of the statement. 
\end{enumerate}
\label{def:zk-snark}
\end{definition}
A witness is usually an auxiliary piece of information necessary to convince the verifier of the truth a statement. In manufacturing, the witness is the correct sequence of state transitions, i.e., the correct specification that the manufacturer must know. 

For implementation, we use gnark, an open source Go library for zk-SNARKs that exposes a high-level API for programming arithmetic circuits \cite{gnark}. Gnark contains sub-libraries that handle the frontend and the backend phases of constructing arithmetic circuits. The frontend library provides an API for describing mathematical statements that can be converted into arithmetic circuits. The backend APIs compile the objects created by the frontend APIs into two particular zk-SNARK constructions: R1CS \cite{groth2016} and PlonK \cite{plonk}. We use the R1CS constructions for our zk-SNARKs due to its superior performance, evidenced by its widespread use.
One useful design of gnark is that the library is exposed as an API. This is crucial for tightly integrating the verifiable proof logic with the chaincode infrastructure since the Golang toolchain only needs to import the gnark library source files and no additional resources outside the Golang toolchain, such as shared libraries. Any Go objects resulting from the verifiable proof logic can be used directly in the chaincode.
The verifiable proofs of state sequences are implemented in chaincodes using gnark are described as follows. 

\subsubsection{Circuit Compilation}

First, the set-membership logical test for the state sequences and auxiliary data must be implemented as arithmetic circuits. This is a three-step process in gnark. The first step consists of implementing a struct whose member variables correspond to the states of the state sequence. The member variables must have the type \texttt{frontend.Variable} in order to be visible to gnark. The second step consists of defining a series of constraints using the variables defined in the previous step. In gnark, this must be done by providing an implementation of the receiver function \textsf{Define} over the struct defined in the previous step. The logic describing the set-membership logical test is implemented within the body of \textsf{Define}. We note that \textsf{Define} 
is not explicitly used within the codebase, rather, it is used by the gnark library internally. Finally, the circuit is compiled by invoking the function \textsf{Compile}, which takes as its input an instance of the circuit, the choice of elliptic curve, and the choice of zk-SNARK backend and returns a compiled constraint system.

\subsubsection{Key Setup}

After the set-membership logical test is implemented, the corresponding keys necessary for the proof generation and verification needs to be generated. This is done in gnark by calling the function \textsf{Setup} which takes as input a compiled constraint system and returns an evaluation key and a verification key. 

\subsubsection{Prove}

The final step is in generating the actual verifiable proof. This is done in gnark by calling the function \textsf{Prove}, which takes in the compiled constraint system, the private key and a witness containing the expected state sequence and returns a proof. 
After the verifiable proof is generated, it must be stored locally on each full node in the blockchain network. 
While the chaincode API does not allow, by default, storing data local to a chaincode, one workaround is to exploit the fact that chaincodes run in their own VMs (Docker container) and serialize the verifiable proof and the verification key onto both the local filesystem within the chaincode VM and the blockchain. 

\subsubsection{Verify} When the verification contract is invoked, it loads the verifiable proof and key from the local storage, verifies the state sequence and sends the result onto the orderer.

\subsection{Blockchain Network Architecture}

Permissioned blockchains mainly differ from unpermissioned blockchains in that permissioned blockchains require participants to be trusted before they are allowed to participate in the network. This model of trust is identical to that of industrial settings and for this reason permissioned blockchains are well-suited for implementing a factory automation blockchain. 

We implement the permissioned blockchain network using Hyperledger Fabric \cite{hlf2018}, a Docker-based permissioned blockchain framework written in Go. Nodes in the network are Docker containers that communicate with each other via conventional internet protocols. The Fabric library compiles into two main binaries: the \textsf{orderer} and the \textsf{peer} binaries. The \textsf{orderer} binary contains the ordering service, whereas the \textsf{peer} binary contains the functionality to run a full node. Blockchain functionality 
is implemented by providing the functions from the chaincode interface in Fabric. Chaincodes are compiled into binaries which are then run on peers. The separation of the peer and chaincode executables isolates the node from any errors in the smart contract.

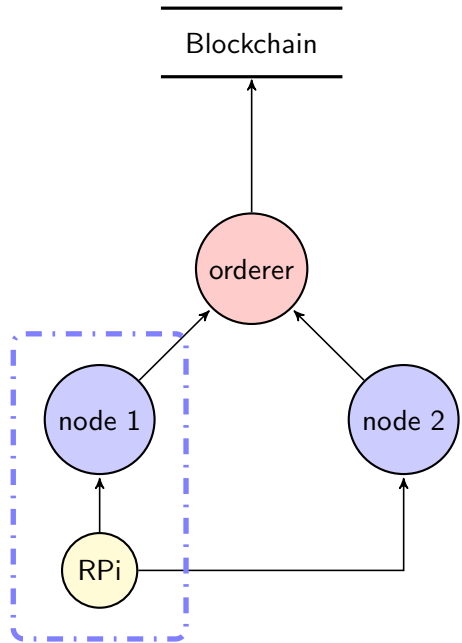
\begin{figure}[t]
\centering
\begin{tikzpicture}[
  font=\sffamily,
  every matrix/.style={ampersand replacement=\&,column sep=2cm,row sep=2cm},
  source/.style={draw,thick,rounded corners,fill=green!20,inner sep=.3cm},
  process/.style={draw,thick,circle,fill=blue!20},
  client/.style={draw,thick,circle,fill=yellow!20},
  orderer/.style={draw,thick,circle,fill=red!20},
  sink/.style={source,fill=green!20},
  choice/.style={source,fill=white!20},
  datastore/.style={draw,very thick,shape=datastore,inner sep=.3cm},
  dots/.style={gray,scale=2},
  to/.style={->,>=stealth',shorten >=1pt,semithick,font=\sffamily\footnotesize},
  every node/.style={align=center}]

\node[datastore] (bcnetwork) at (0,5) {Blockchain};
\node[orderer] (orderer) at (0,2) {orderer};
\node[process] (node1) at (-2,0) {node 1};
\node[process] (node2) at (2,0) {node 2};
\node[client] (rpi) at (-2,-2) {RPi};

\draw[to] (rpi) -- (node1);
\draw[to] (rpi) -- (2,-2) -- (node2);

\draw[to] (node1) -- (orderer);
\draw[to] (node2) -- (orderer);

\draw[to] (orderer) -- (bcnetwork);





\tikzset{blue dotted/.style={draw=blue!50!white, line width=2pt,
                           dash pattern=on 1pt off 4pt on 6pt off 4pt,
                            inner sep=4mm, rectangle, rounded corners}};

                        
                        
\node (first dotted box) [blue dotted,
                        fit = (rpi) (node1) ] {};

                            
\end{tikzpicture}
\caption{PoC blockchain network architecture}
\label{fig:blockchain-arch}
\end{figure}

The architecture of the blockchain network is illustrated in Fig.~\ref{fig:blockchain-arch} and consists of an orderer, two full nodes, and a Raspberry Pi (RPi). The dashed blue box signifies the fact that the full node and the RPi (PLC) belong to the same organization. We describe each in the following.

\subsubsection{Full nodes}

There are two full nodes in the factory automation blockchain network represented in purple in Fig.~\ref{fig:blockchain-arch}. Full nodes store a full copy of the blockchain. More importantly, each full node stores a copy of, compiles, and executes verification contracts. The results from each full node are sent to the orderer for consensus. 

Full nodes represent participants in a factory automation network that need to ensure that an industrial process was executed according to specification. For example, a full node can be run within a factory as an edge node that ensures that internal quality controls are met. Another example of a participant that may run a node is a downstream producer that needs to ensure that incoming products are indeed manufactured according to specification. 

\subsubsection{Orderer}

In permissoned blockchains, an orderer node orders the transactions in the network, thereby providing consensus. In the PoC, the orderer node is implemented using the conventional application of Fabric libraries. 

\subsubsection{PLC}

The PLC is implemented using a Raspberry Pi. It controls field devices and invokes the verification contracts. Its implementation details have already been discussed. 

\subsection{Implementation and Discussion}

In the PoC, the process consists of identifying an incoming part by scanning its barcode, and machining the part in a CNC machine tool using  a specific input file and machine configuration. The Raspberry Pi runs 64 bit Ubuntu 20.04 emulated the functionality of a PLC. The barcode scanner and  CNC milling machine are connected to the PLC and act as field devices. 
During machining, 
the PLC uses gpio pins to monitor several status flags provided by the machine tool such that faults in the machining process could be detected.

\begin{figure}[t]
  \includegraphics[width=\linewidth]{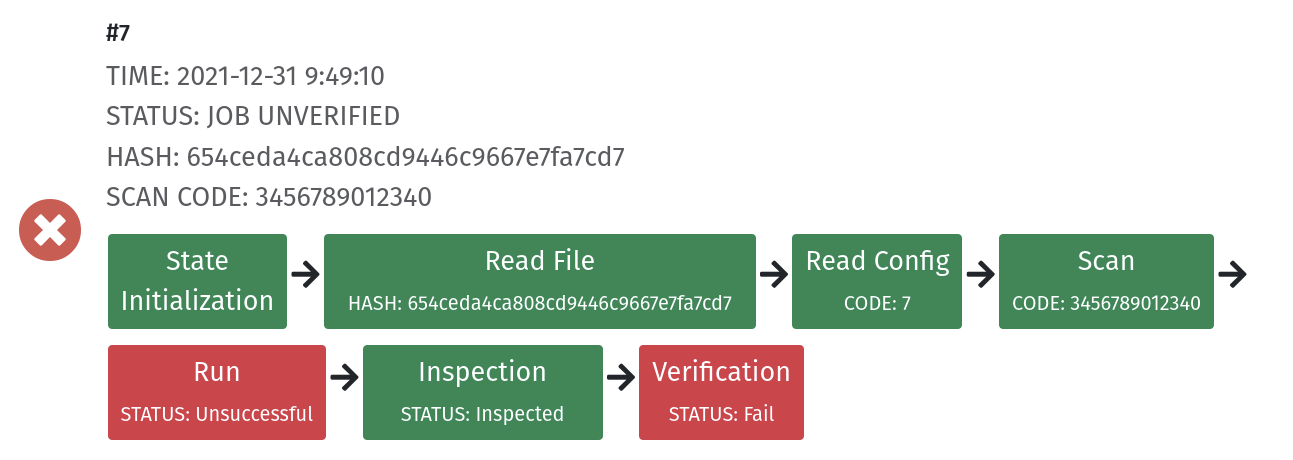}
  \hspace{1pt}
  \includegraphics[width=\linewidth]{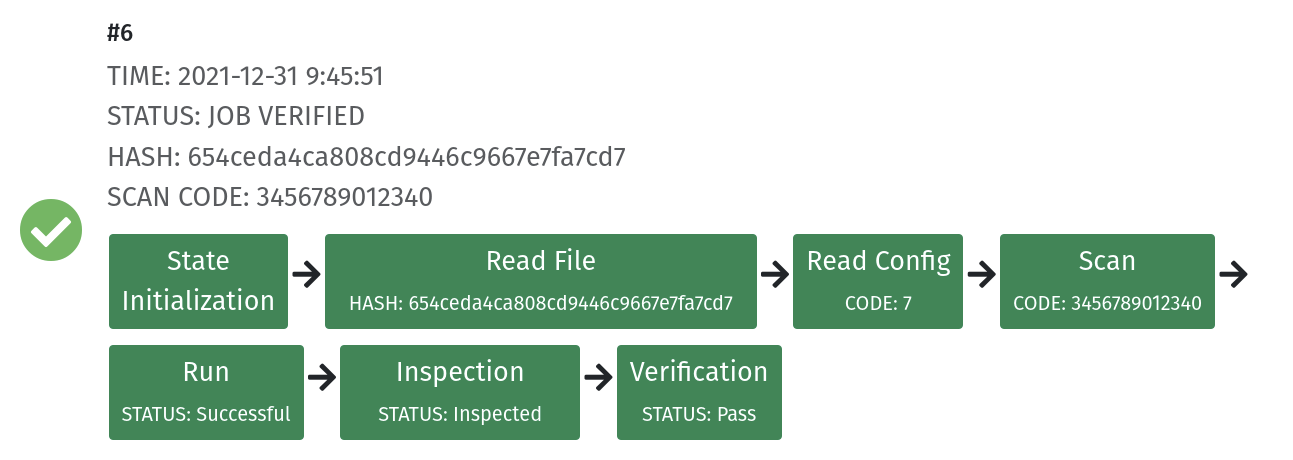}
  \caption{PoC verification results recorded on the blockchain}
  \label{fig:poc_result}
\end{figure}

An application developed in C emulates a PLC program implementing a finite state machine (FSM) for sequencing individual steps of the physical manufacturing process. Valid state sequences derived from the FSM were used to generate verification proofs, the results of which are presented in Fig.~\ref{fig:poc_result}.
The state sequence consists of the following states: an initial state, a read-file state with an expected file hash, a read-config state with an expected success code, a scan state, an inspection state, and an end state with a verification result code.

In the figure, one can see both a failed run and a successful run. In the first case, verification results in a failed state sequence; in the second, the state sequence is successfully verified. In the graphic, red boxes indicate where in the sequence (finite state machine) failures have been encountered. 
The prototype demonstrates that unexpected changes in manufacturing processes, 
previously difficult to verify, 
can now be verified and memorialized in a trusted manner using blockchain.
The PoC therefore shows that blockchain can make verifiable computing practical for physical processes within a manufacturing setting, illustrating that physical processes can be made verifiable through blockchain-based verifiability.

In this work, we determined the state sequences corresponding to specific outcomes manually. This was only possible because the physical process was very simple and computing the set of valid state sequences for an arbitrary PLC program controlling a complex process was outside the scope of this work. 
As discussed previously, this work assumes that 
control and field devices are trusted; future work will work to enable verifiability at this level. 

%
%

\section{Conclusion}
\label{sec:concl}

This work considered the application of a blockchain-based verifiable computing scheme to physical manufacturing processes. We adapted verifiable computing, developed for abstract mathematical objects, to physical processes by establishing a one-to-one relationship between physical transformations carried out by field devices to the state sequences that direct them on a PLC. The verifiability of physical processes in and of itself is not enough to enable the verifiability of manufacturing processes since a direct application of verifiability would be computationally intractable. To this end, we introduced a blockchain-based solution consisting of embedding verifiable proofs in smart contracts. The public nature of smart contracts within blockchain networks makes verifiable proofs trusted, enabling reuse. The underlying blockchain acts an immutable record of the verification of manufactured items that have undergone some physical transformation. We implemented a proof of concept based on Hyperledger Fabric demonstrating the viability of our scheme. 

\bibliographystyle{IEEEtran}
\bibliography{main}

\end{document}